# Imaging low-mass planets within the habitable zone of α Centauri

**Authors:** K. Wagner,[1, 2, *] A. Boehle,[3] P. Pathak,[4] M. Kasper,[4] R. Arsenault,[4] G. Jakob,[4] U. Käufl,[4] S. Leveratto,[4] A.-L. Maire,[5] E. Pantin,[6] R. Siebenmorgen,[4] G. Zins,[4] O. Absil,[5] N. Ageorges,[7] D. Apai,[1,2,8] A. Carlotti,[9] É. Choquet,[10] C. Delacroix,[5] K. Dohlen,[10] P. Duhoux,[4] P. Forsberg,[11] E. Fuenteseca,[4] S. Gutruf,[4] O. Guyon,[1, 12, 13] E. Huby,[15] D. Kampf,[7] M. Karlsson,[11] P. Kervella,[15] J.-P. Kirchbauer,[4] P. Klupar,[13] J. Kolb,[4] D. Mawet,[16] M. N'Diaye,[17] G. Orban de Xivry,[5] S. P. Quanz,[3] A. Reutlinger,[7] G. Ruane,[16, 18] M. Riquelme,[4] C. Soenke,[4] M. Sterzik,[4] A. Vigan,[10] and T. de Zeeuw[4, 19, 20]

[1] Dept. of Astronomy and Steward Observatory, University of Arizona, 933 N Cherry Ave, Tucson, AZ 85719, USA.

[2] NASA Nexus for Exoplanet System Science, Earths in Other Solar Systems Team.

[3] Institute for Particle Physics and Astrophysics, ETH Zurich, Wolfgang-Pauli-Strasse 27, 8093 Zürich, Switzerland.

[4] European Southern Observatory, Karl-Schwarzschild-Straße 2, 85748 Garching bei München, Germany.

[5] STAR Institute, Université de Liège, Allée du Six Août 19c, B-4000 Liège, Belgium.

[6] AIM, CEA, CNRS, Université Paris-Saclay, Université Paris Diderot, Sorbonne Paris Cité, F-91191 Gif-sur-Yvette, France.

[7] Kampf Telescope Optics, Alois-Gilg-Weg 7, 81373 München, Germany.

[8] Lunar and Planetary Laboratory, University of Arizona, 1629 E University Blvd, Tucson, AZ 85721, USA.

[9] Univ. Grenoble Alpes, CNRS, IPAG, 38000 Grenoble, France.

[10] Aix Marseille Univ, CNRS, CNES, LAM, Marseille, France.

[11] Department of Materials Science and Engineering, Ångström Laboratory, Uppsala University, Uppsala, Sweden.

[12] Subaru Telescope, 650 N Aohoku Pl, Hilo, HI 96720, USA.

[13] The Breakthrough Initiatives, NASA Research Park, Bld. 18, Moffett Field, CA, 94035, USA.

[14] James C. Wyant College of Optical Sciences, University of Arizona, Tucson, AZ, USA

[15] LESIA, Observatoire de Paris, 5 Place Jules Janssen, 92190 Meudon, France.

[16] California Institute of Technology, 1200 E. California Blvd., Pasadena, CA 91125, USA.

[17] Université Côte d'Azur, Observatoire de la Côte d'Azur, CNRS, Laboratoire Lagrange, Nice, France.

[18] Jet Propulsion Laboratory, California Institute of Technology, 4800 Oak Grove Dr., Pasadena, CA 91109, USA.

[19] Sterrewacht Leiden, Leiden University, Postbus 9513, 2300 RA Leiden, The Netherlands.

[20] Max Planck Institute for Extraterrestrial Physics, Giessenbachstraße 1, 85748 Garching, Germany.

* Correspondence to: kevinwagner@email.arizona.edu

**Abstract:** Giant exoplanets on wide orbits have been directly imaged around young stars. If the thermal background in the mid-infrared can be mitigated, then exoplanets with lower masses can also be imaged. Here we present a ground-based mid-infrared observing approach that enables imaging low-mass temperate exoplanets around nearby stars, and in particular within the closest stellar system, α Centauri. Based on 75–80% of the best quality images from 100 hours of cumulative observations, we demonstrate sensitivity to warm sub-Neptune-sized planets throughout much of the habitable zone of α Centauri A. This is an order of magnitude more sensitive than state-of-the-art exoplanet imaging mass detection limits. We also discuss a possible exoplanet or exozodiacal disk detection around α Centauri A. However, an instrumental artifact of unknown origin cannot be ruled out. These results demonstrate the feasibility of imaging rocky habitable-zone exoplanets with current and upcoming telescopes.

**Introduction:** A primary pursuit of modern astronomy is the search for worlds that are potentially similar to Earth. Such worlds would help us to understand the context of our own planet and would themselves become targets of searches for life beyond the solar system (e.g., *1–3*). Meanwhile, giant exoplanets have been imaged on wide orbits–enabling direct studies of their orbits and atmospheres (e.g., *4–6*). To enable finding and exploring potentially Earth-like planets, exoplanet imaging capabilities are progressing towards lower-mass planets in the habitable zones of nearby stars (e.g., *7–9*). In this context, habitable refers to the possibility of a planet with a broadly Earth-like atmosphere to host liquid water on it surface.

The nearest stellar system, α Centauri, is among the best-suited for imaging habitable-zone exoplanets (e.g., *10–12*). The primary components α Centauri A and B are similar in mass and temperature to the Sun, and their habitable zones are at separations of about one au (see *13* and Fig. 1). At the system's distance of 1.3 pc, these physical separations correspond to angular separations of about one arcsecond, which can be resolved with existing 8-m-class telescopes. However, no planets are currently known to orbit either star. Measurements of the stars' radial velocity (RV) trends (*14*) exclude planets more massive than $M\sin i \geq 53$ Earth-masses ($M_\oplus$) in the habitable zone of α Centauri A, and $\geq 8.4$ $M_\oplus$ for α Centauri B. Lower-mass planets could still be present and dynamically stable (e.g., *15*). The tertiary M-dwarf component of the system, Proxima Centauri, also hosts at least two planets more massive than Earth (*16, 17*) that were discovered through the star's RV variations.

Conventional exoplanet imaging studies (e.g., *18–20*) have operated at wavelengths of $\lambda \lesssim 5$ μm, in which the background noise is relatively low (i.e., the sensitivity is dominated by residual starlight), but in which temperate planets are faint compared to their peak emission in the mid-infrared ($\lambda \sim 10$–20 μm). The exoplanets that have been imaged are young super-Jovian planets on wide orbits ($a > 10$ au) with temperatures of $\sim 10^3$ K (e.g., *18–21*). Their high temperatures are a remnant of formation and reflect their youth ($\sim 1$–100 Myr, compared to the Gyr ages of typical stars). Imaging potentially habitable planets will require imaging colder exoplanets on shorter orbits around mature stars. This leads to an opportunity in the mid-infrared ($\sim 10$ μm), in which temperate planets are brightest. However, mid-infrared imaging introduces significant challenges. These are primarily related to the much higher thermal background–that saturates even sub-second exposures–and also the $\sim 2$–5× coarser spatial resolution due to the diffraction limit scaling with wavelength. With current state-of-the-art telescopes, mid-infrared imaging can resolve the habitable zones of roughly a dozen nearby stars, but it remains to be shown whether sensitivity to detect low-mass planets can be achieved.

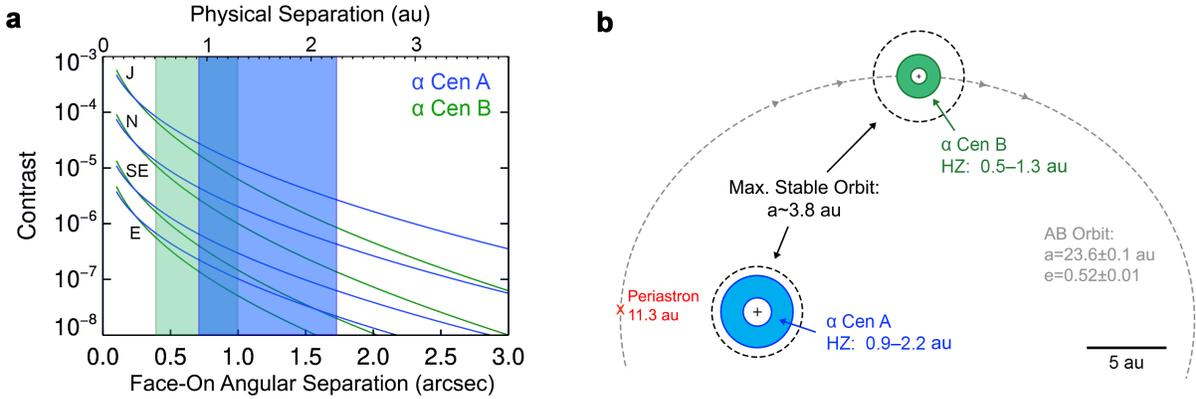

**Fig. 1. Simulated planet brightness and basic properties of the α Centauri system. a** *N*-band (10–12.5 µm) contrast vs. angular separation of planets around α Centauri A (blue) and B (green), assuming face-on circular orbits, a Bond albedo of 0.3 and internal heating that provides an additional 10% of the planets' equilibrium temperatures. The curves correspond from bottom to top to planetary radii equivalent to that of Earth, a Super-Earth (1.7×Earth's radius, R⊕), Neptune, and Jupiter. The blue and green shaded regions show the location of the classical habitable zones around α Centauri A and B, respectively (*13*). **b** Diagram of the orbital properties and approximate habitable zones of the α Centauri AB system. Note that this diagram does not show the 79º inclination of the orbit as seen from Earth, or the tertiary dwarf star, Proxima Centauri, at ~10⁴ au.

In this work, we present the results of the New Earths in the α Centauri Region (NEAR: *22, 23*) experiment. As part of Breakthrough Watch (*24*), NEAR aims to demonstrate experimental technologies and techniques to facilitate directly imaging low-mass habitable-zone exoplanets. Specifically, NEAR aims to demonstrate that low-mass exoplanets can be imaged in a practical, but unprecedented amount of observing time (~100 hr) by conducting a direct imaging search for habitable-zone exoplanets within the nearest stellar system, α Centauri. We describe the NEAR campaign, an analysis of the sensitivity to habitable-zone exoplanets, and an assessment of a candidate detection. Finally, we discuss possibilities for imaging rocky habitable-zone exoplanets around α Centauri and other nearby stars with these techniques.

**Results:** NEAR explored an instrumental setup and observing strategy designed to push the capabilities of ground-based exoplanet imaging toward mid-infrared wavelengths of ~10 µm. The mid-infrared camera (VISIR: *25*) on the Very Large Telescope (VLT) was upgraded for NEAR to implement several new technologies, such as a mid-infrared optimized annular groove phase mask (AGPM: *26*) coronagraph (*27, 28*) and shaped-pupil mask (*29, 30*) to suppress the starlight. VISIR was moved to unit telescope 4 (UT4/Yepun) of the VLT, which is equipped with a deformable secondary mirror (DSM, *31*). The DSM is a crucial element of NEAR's strategy, as it enabled performing adaptive optics (AO) without additional non-cryogenic corrective optics whose thermal emission would contribute to the total background. The DSM was also utilized to alternate the position of the binary behind the coronagraph (chopping) with a frequency of ~10 Hz, while pausing the AO state during the transition between stars. Subtracting the images taken during alternate positions removed much of the background–partly including the contribution from the annular groove phase mask (*32*). With this strategy, planets around either star would appear centered on the coronagraph, with planets orbiting α Centauri A appearing as positive point sources and planets orbiting α Centauri B appearing as negative sources. Additional details about the experiment design can be found in (*22, 23*, and methods, instrumental setup and observing Strategy).

α Centauri was observed from 2019 May 23 to 2019 June 11. The nightly conditions can be found in Supplementary Table 1. An additional night of data was taken on 2019 June 27. Enough time separates the collection of these data from the initial observations such that orbital motion complicates combining these with the rest of the data (*33, 34*). However, this extra night provides a useful astrometric check for brighter planet candidates. Including these data, we collected a cumulative exposure time of approximately 100 hours, out of which 23 hours were not used because of mediocre data quality as a result of high sky background, coronagraph misalignment, or AO problems. The remaining 76.9 hours of good quality data were used for the subsequent analysis. Further details can be found in methods, data reduction and processing.

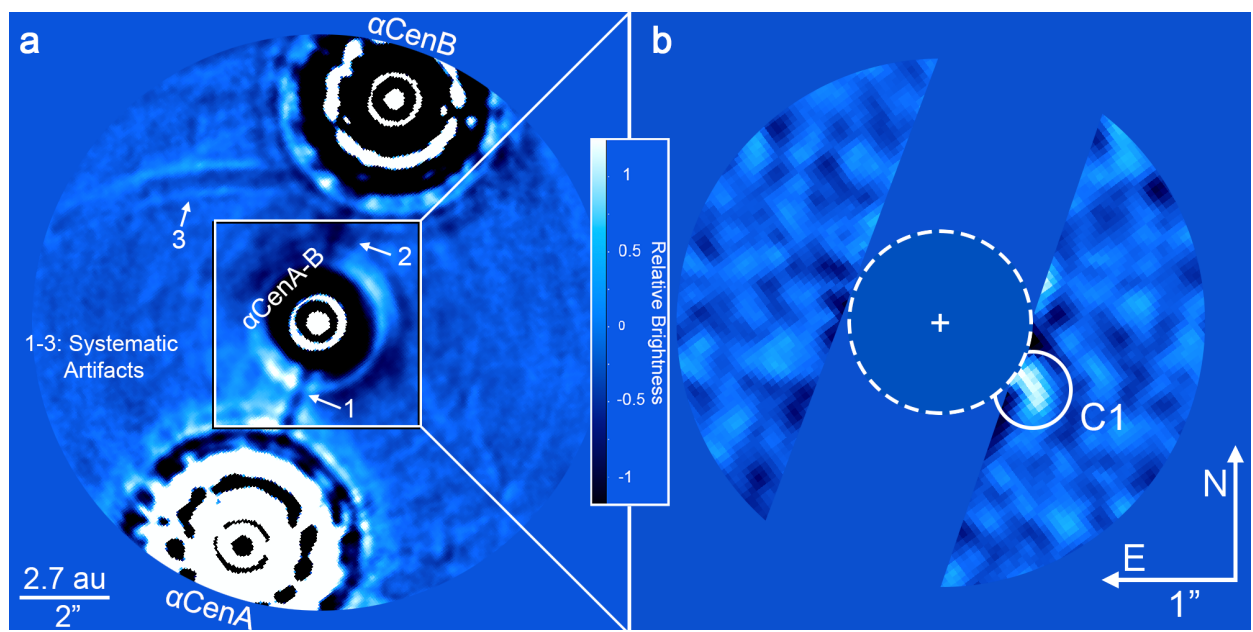

**Fig. 2**. **Mid-IR Images of α Centauri. a** high-pass filtered image without PSF subtraction or artifact removal. The α Centauri B on-coronagraph images have been subtracted from the α Centauri A on-coronagraph images, resulting in a central residual and two off-axis PSFs to the SE and NW of α Centauri A and B, respectively. Systematic artifacts labeled 1–3 correspond to detector persistence from α Centauri A, α Centauri B, and an optical ghost of α Centauri A. **b** Zoom-in on the inner regions following artifact removal and PSF subtraction. Regions impacted by detector persistence are masked for clarity. The approximate inner edge of the habitable zone of α Centauri A (*13*) is indicated by the dashed circle. A candidate detection is labeled as 'C1'.

Following stellar point spread function (PSF) subtraction, the brightest features in the images are systematic artifacts (see Fig. 2 & Supplementary Fig. 1). The most significant artifact is detector persistence accumulated during the chopping sequence. The second most significant artifacts that appear in the final images are negative arcs due to optical ghosts (reflections) of the off-axis PSF of α Centauri A that are introduced by the dichroic beam-splitter and spectral filter. These artifacts limit the overall image sensitivity and increase the false positive probability within specific regions. To improve the overall image sensitivity, we modeled and subtracted each of the known artifacts (see Supplementary Fig. 1). The full-frame image prior to artifact subtraction and a zoom-in on the habitable zone following artifact subtraction are shown in Fig. 2.

An Earth-sized planet at a separation of $\rho \sim 1.2''$ ($a \sim 1.5$ au) around $\alpha$ Centauri A would appear with $\sim 10^{-7}$ contrast at $\lambda \sim 10$ μm (see Fig. 1, and also *22*). A super-Earth ($R \sim 1.7$ R$_\oplus$) at the inner edge of the habitable zone ($\rho \sim 0.8''$) would appear with a contrast of $\sim 10^{-6}$. As an assessment of the detector's fundamental sensitivity limit (in the absence of residual stellar flux and spatially correlated noise), we examined the standard deviation of pixel intensities within 1.2 $\lambda/D$ ($\sim 0.35$ arcsec, or $\sim 8$ pixels) in diameter in a region of the detector far from $\alpha$ Centauri A and B (see supplementary methods, background-limited sensitivity). We found this value (multiplied by the square root of the number of pixels contained within the aperture) to be $\sim 1.67 \times 10^{-7}$ contrast with respect to $\alpha$ Centauri A, or about $\sim 22$ μJy. The pixel-to-pixel noise increases toward the glow of the AGPM (*32*). At 1" separation, the standard deviation of pixel intensities is roughly doubled by the glow.

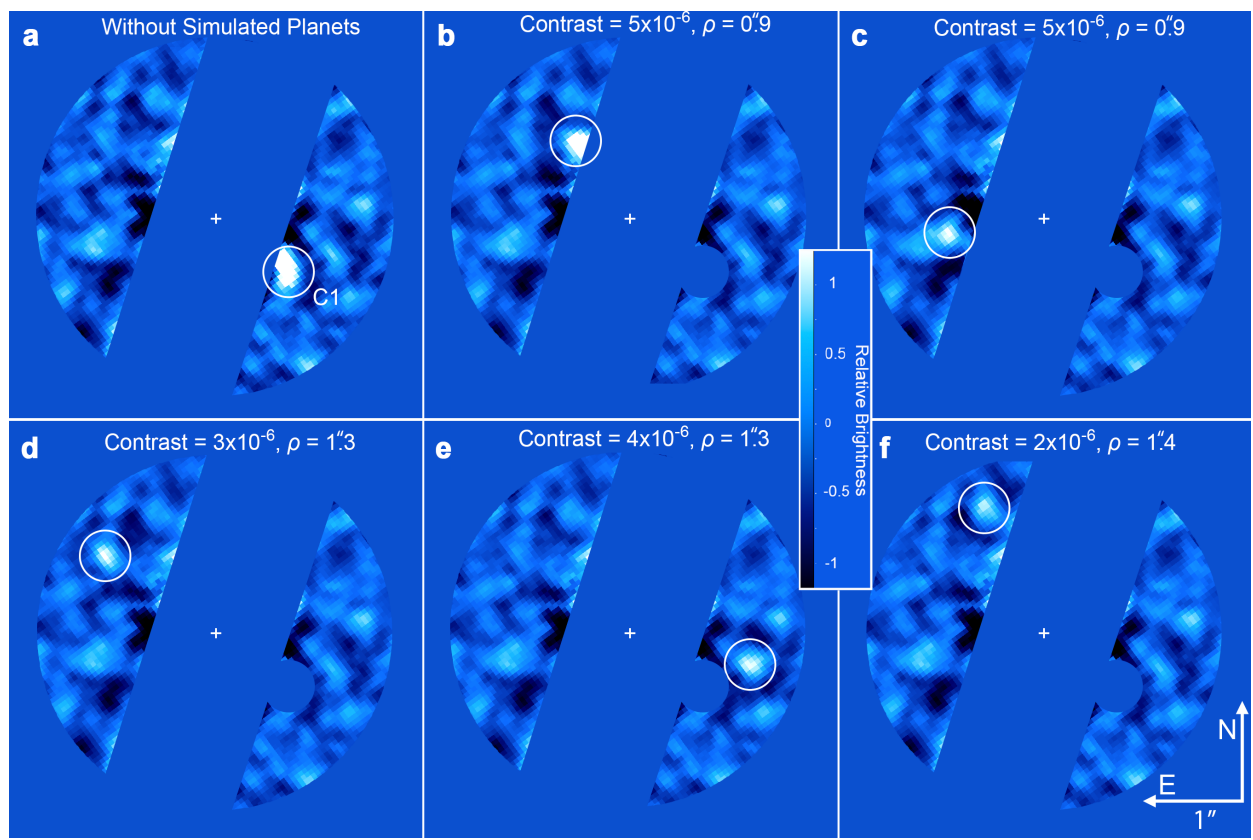

**Fig. 3**. **NEAR campaign image (a) and those with simulated planets (b-f).** Each image has been PSF-subtracted following removal of known artifacts. The location of C1 has been masked in **b-f** so that the simulated planets (indicated in these panels by white circles) can be clearly identified. These examples demonstrate the lower brightness limit at which simulated planets are identifiable. The bottom right panel represents the limiting case at which the source is marginally identifiable among speckles of similar brightness.

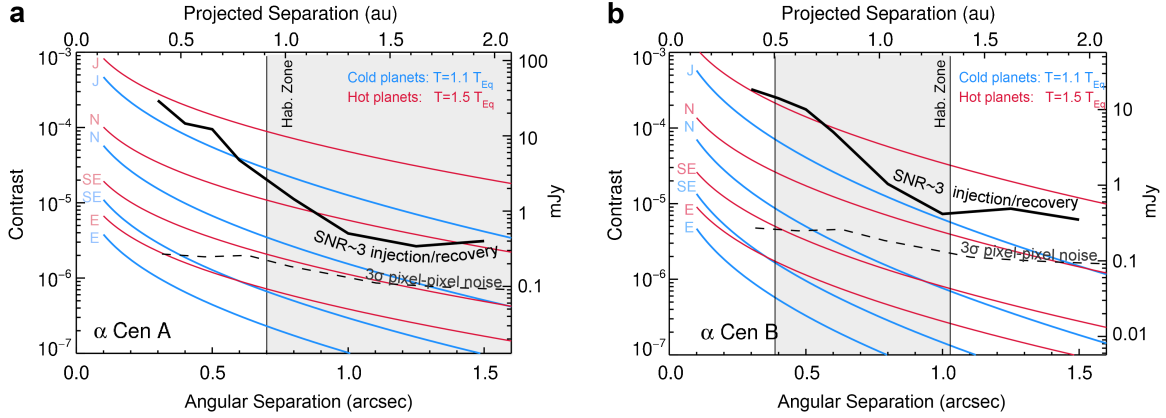

**Fig. 4. Sensitivity of the NEAR data.** Results for α Centauri A & B are shown in **a** and **b**, respectively. The bold curve shows the sensitivity to point sources computed from simulated injection and recovery tests and the dashed curve shows the background noise contribution from the variation of pixel-to-pixel intensities. The red and blue curves represent simulated planets of equivalent radii to those in the solar system, with the addition of an $R$=1.7$R_\oplus$ Super-Earth (SE). Each model planet's temperature is set by the assumption of thermal equilibrium at a given separation with an $A_B$=0.3 Bond albedo and internal heating included as 10% or 50% of the equilibrium temperature; similar to conditions of the solar system planets (*46, 47*).

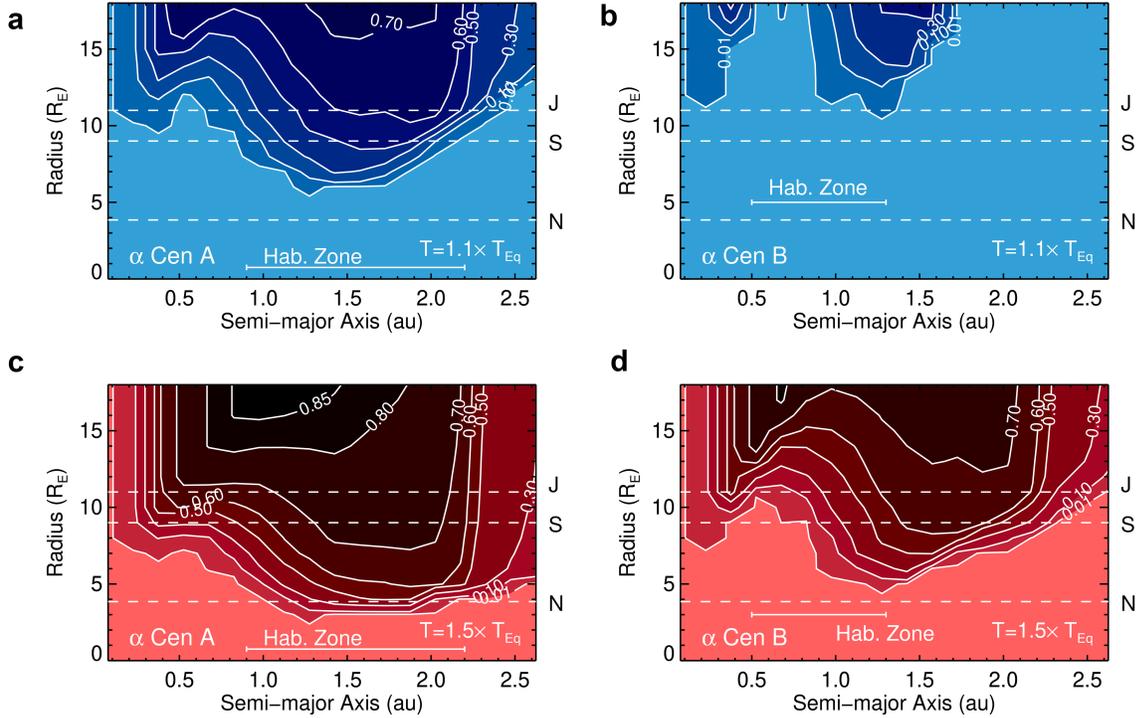

**Fig. 5. Completeness to planets of various radii and orbital semi-major axes. a** and **b** assume $A_B$=0.3 and internal heating contributing 10% of the equilibrium temperature, whereas **c** and **d** assume $A_B$=0.3 and internal heating contributing 50% of the equilibrium temperature. Radius and semi-major axis were uniformly sampled along with an inclination prior of P($i$) ∝ sin $i$ (see supplementary mfethods, completeness analysis). The dashed lines correspond to the radii of Neptune, Saturn, and Jupiter (N, S, and J, respectively). Contour units are normalized.

For an empirical assessment of the detection sensitivity to planets (i.e., point sources), we performed numerous simulated point-source injection and retrieval tests via forward modelling injected signals of planetary-brightness throughout the data processing (see supplementary methods, simulated planet injection and retrieval tests). Injected sources are identified in the images at about an order of magnitude above the (1-σ) background noise, or ~2–3×10⁻⁶ contrast to α Centauri A (see Fig. 3–4). Over much of the habitable zone of α Centauri A, the image sensitivity is sufficient to detect $R\geq 9$ R$_\oplus$ planets in radiative thermal equilibrium (i.e., Saturn-sized, with $T\sim 300$ K), and smaller planets with additional heat or low Bond albedos. Within α Centauri B's habitable zone, the images reach sensitivities to detect Jupiter-sized planets ($R\sim 11$ R$_\oplus$) with a small amount of additional heat ($T=1.1\times T_{eq}$) or a low Bond albedo.

We converted the sensitivity analysis into a completion estimate using a Monte Carlo simulation to draw randomly sampled orbital parameters (see Fig. 5 and supplementary methods, completeness analysis for details). With no prior orbital constraints, the NEAR data reach a maximum completeness of ~80% for Jovian-sized planets, or ~85% for slightly larger (i.e., inflated) planets. The maximum completeness is less than unity since the projected separation can be smaller than a given semi-major axis, or behind the persistence stripes. At the extreme end of the detection limits, there is a ~1–10% chance of detecting a warm $R\sim 3$ R$_\oplus$ planet orbiting α Centauri A. For comparison, the plausible radii range of rocky exoplanets extends up to $R\sim 1.75$ R$_\oplus$ (*35, 36*).

**Discussion:** The primary goal of the NEAR campaign is to demonstrate the capabilities of thermal infrared exoplanet imaging. The results showed that the sensitivity is background limited and follows a signal to noise ratio (SNR) $\propto \sqrt{t}$ relation in image regions far from the center ($\geq 7$ $\lambda/D$). The achieved sensitivity in such regions in one hour of observations (5σ) is ~0.75 mJy (see *37*, and supplementary methods, background-limited sensitivity). The habitable zone of α Centauri A is located at ~1" (see Figure 1 and *13*), which corresponds to the contrast-limited region of ~3.5 $\lambda/D$ in the 10–12.5 μm bandpass for the 8.2-m VLT. With a 39-m Extremely Large Telescope (ELT), ~1" would correspond to ~17.5 $\lambda/D$ and would therefore likely be close to background-limited. In that case, the SNR scales $\propto \sqrt{t}\,D^2$, and thus the time required to reach a given SNR scales $\propto D^{-4}$. The predicted sensitivity at 1" of a NEAR-like instrument on an ELT would therefore be ~35 μJy (5σ in one hour). This would in principle be sufficient to detect an Earth-analogue planet around α Centauri A (~ 20 μJy) in just a few hours, which is consistent with expectations for the ELTs (*7, 38*).

If the ELT's performance at 1" is instead contrast-limited, then Earth-like planets could still be imaged, since the intensity of quasi-static speckles produced by optical polishing errors at a given angular separation scales $\propto D^{-2}$ or steeper (*39*). The NEAR campaign demonstrated a final contrast-limited sensitivity (SNR~3 in 77 hours) of ~3×10⁻⁶ contrast to α Centauri A (~0.4 mJy) at ~1". Extrapolating to the larger aperture of the ELT suggests a contrast limit of ~1.5×10⁻⁷ or better at 1". This again supports the predictions that the ELTs will reach sensitivity levels sufficient to image Earth-analogue planets around α Centauri A (*7, 38*). These estimates may also be expected to be improved, since the increased local background produced near the center due to the glow of the AGPM can be mitigated by a cold pupil stop in front of the coronagraphic mask, as implemented by the current instrument design plans for the METIS instrument (*40*). The contrast-limited performance of future instruments could also be improved by pupil apodizers (*41*) and non-common path aberration calibration mechanisms (*42*) that were not available for the design of NEAR. Improvements to mid-IR detector technologies could also lead to significant improvements.

A secondary goal of NEAR is to explore the habitable zones of α Centauri. While designed to detect thermal emission from exoplanets, our observations could also detect warm exozodiacal dust (e.g., *43*). Here, we consider whether either such detection is present in the images. In a relatively clean region of the image, there is one point-like feature (signal to noise ratio ~ 3) that is not associated with any known detector artifacts. We refer to this source as Candidate 1, or C1 (see Fig. 2). C1 appears with a brightness that would be expected of a giant planet ($R$~3–11 R$_\oplus$) at ~1.1 au from α Centauri A and with an elongation on the order of ~0.1 arcsec–consistent with orbital motion of a planet in an $i$~70º orbit throughout the nineteen-day campaign. Notably, the detection of C1 is repeatable in multiple independent subsets of the data (see Supplementary Fig. 2), which makes it unlikely to be a random false positive. Based on pre-imaging conducted a decade prior to the NEAR campaign, we can exclude the possibility that C1 is a background source (see *44* and supplementary methods, pre-imaging for background sources). Therefore, we consider C1 to be a plausible exoplanet and/or exozodiacal disk candidate. While C1 cannot be explained by presently known systematic artifacts, an independent experiment is necessary to exclude this third possibility.

RV observations exclude the presence of $M\sin i \geq 53$ M$_\oplus$ planets within the habitable zone α Centauri A (*14*). Assuming $R \propto M^{0.55}$ (*45*), this limit corresponds to $R < 7$ R$_\oplus$. Among a range of radii of $R$~3.3−7 R$_\oplus$, the brightness of C1 can be explained with a level of additional heating sufficient to raise the planet's temperature by 5−50% of its radiative equilibrium temperature (assuming $A$=0.3). The lower limit is motivated by Neptune's effective temperature, which is ~50% higher than its radiative equilibrium temperature (*46, 47*). C1 could also be an exozodiacal disk with ~60 zodis of dust, and with a stellocentric offset of ~0.3 au to the SW (see supplementary methods, exozodiacal dust disk modeling). This would be a relatively large dust mass for a G-type star (*48*), but would be within precedent (e.g., ε Eridani has ~200 zodis: *48*). This dust mass is also consistent with the upper limits from the far-IR spectrum of α Centauri A (≤100 zodis, *49*). In other words, C1 is not a known systematic artifact, and is consistent with being either a Neptune-to-Saturn-sized planet or an exozodiacal dust disk.

The habitable zones of α Centauri and other nearby stars could host multiple rocky planets–some of which may host suitable conditions for life. With a factor of two improvement in radius sensitivity (or a factor of four in brightness), habitable-zone super-Earths could be directly imaged within α Centauri. An independent experiment (e.g., a second mid-infrared imaging campaign, as well as RV, astrometry, or reflected light observations) could also clarify the nature of C1 as an exoplanet, exozodiacal disk, or instrumental artifact. If confirmed as a planet or disk, C1 would have implications for the presence of other habitable zone planets. Mid-infrared imaging of the habitable zones of other nearby stars, such as ε Eridani, ε Indi, and τ Ceti is also possible. In the next decade, application of these techniques with extremely large telescopes (e.g., with ELT/METIS: *7, 38, 40*) will enable sensitive exploration of the habitable zones of these and other nearby stars.

**Methods:**

Instrumental Setup and Observing Strategy

The VLT Imager and Spectrometer for the Mid-IR (VISIR: *25*) was significantly upgraded for the NEAR experiment. VISIR was coupled with the VLT's DSM (*31*), which enabled the implementation of AO without increasing the number of warm optics that would add to the thermal background. The AO correction resulted in typical Strehl ratios in excess of 97%. The DSM was used for ~8 Hz chopping to enable tracking and subtracting the systematic excess low frequency noise (ELFN) within the Si:As Aquarius detector, which is a major limitation to the sensitivity of mid-infrared imaging (*50, 51*). Downstream of the DSM, the central starlight was reduced by an AGPM coronagraph optimized for performance at mid-infrared wavelengths (*26, 27*) and a shaped-pupil mask (*28–30*) designed specifically to limit the spatial extent of the Airy pattern from the off-axis star. The Lyot stop is manufactured out of chromium directly deposited on the NEAR spectral filter, which transmits light from 10 to 12.5 μm. This yields a full width at half maximum (FWHM) of ~0.28 arcsec, or ~6 pixels. The observations were done in a pupil-stabilized mode by keeping the Cassegrain instrument at a fixed rotation angle. The detector integration time (DIT) was 6 ms, of which eight frames were averaged and two frames were skipped during the chopping transition. Therefore, each chopping half-cycle equated to 60 ms, resulting in a chopping frequency of 8.33 Hz. For the night of 2019 May 24 we used a DIT of 5.5 ms and normalized the images for this night to account for the difference.

Data Reduction and Processing

We reduced the data for each night of the campaign in a uniform manner with two independent pipelines, which we refer to as the primary and secondary data reductions. The secondary reduction does not implement artifact modelling and subtraction. Therefore, in most cases we utilize the higher fidelity images from the primary reduction and utilize the secondary reduction to confirm the general findings of the first.

We begin by describing the primary data reduction pipeline. From each individual frame with α Centauri A behind the coronagraph we subtracted the mean of the two neighboring frames (chop subtraction). No scaling was performed to normalize the PSFs, as the purpose of chop subtraction is primarily to remove the ELFN and residual background structure such as the AGPM glow. The residual coronagraphic PSF is also partially mitigated by chop subtraction. We then coadded each five hundred image cube into a single image with 24 seconds of equivalent exposure time and combined each of these frames into a single data cube per night of observations. We aligned the frames within each cube via the unocculted point spread function (PSF) of α Centauri B, and determined the precise center of the coronagraphic residual of A-B via rotational centering (*52*). We cleaned the frames by rejecting those whose maximum cross-correlation with respect to the mean of the twenty surrounding frames was less than 0.9 (computed over the radial range of 5−45 pixels, or ~0.2−2 arcsec from the center), which resulted in ~10% frame rejection. At this stage, the known detector artifacts were subtracted from the images (see supplementary methods, artifact modeling I & II). We destriped the images along the horizontal and vertical axes by subtracting the mode of each row and column, and high-pass filtered the data by subtracting a version of each frame from itself after smoothing with a 15-pixel running median. We then stacked and averaged the original frames into 360 second images, and processed the data via both classical angular differential imaging (ADI: *53*) and projection onto eigen images via Karhunen–Loève

Image Processing (KLIP: *54*; specifically using the adaptation from *55*), in which we modelled the PSF with four KL-modes in an annulus from 5−45 pixels. One beam diameter corresponds to ∼14º in azimuth at 1 arcsec, which is significantly larger than the 2.2º of smearing introduced in 360 seconds due to the rotation of the sky. Following PSF subtraction, we applied a second high-pass filter with the same settings to reduce the remaining low-frequency spatial variations. We combined the images within each night using a noise-weighted combination for each pixel, (noise-weighed ADI: *56*) and combined the final images from each night with a variance-weighted mean. The images are shown in Figs. 2–3 and Supplementary Figs. 2–3.

For the second data reduction pipeline, we followed a similar procedure with the following exceptions. Various quality criteria were calculated for the individual chopped images, including: AO correction (ratio of flux in an annulus of radii 6–12 pixels to an aperture of 6 pixels radius), coronagraphic leakage (flux in an aperture of 20 pixels) and sky-background variance calculated over small regions near the edge of the frame. 79.3 hours of data remained after removing the inferior images, which is a similar total exposure time as for the data set created by our other pipeline. Then, the images were co-aligned to the center between the off-axis positions of α Cen A and B and then mean-combined to create frames with an equivalent exposure time of 60 seconds. We then calculated an ADI-based principal component analysis (PCA) model (e.g., *57*) over an annular region around the image center and used this model to subtract the PSF for each observing night separately. Using fake planet injection tests, we optimized the PCA parameters (inner and outer radius of the annulus, number of principal components) to maximize the contrast sensitivity. We arrived at using 15 principal components, an inner radius of 8 px, and an outer radius of 16 px (although the regions further out are also processed). We verified that our conclusions are robust over a range of a few principal components to the maximum number of frames. The final image quality is quite robust with respect to the selected optimization range, and variations of 50% of each parameter do not significantly affect the results. Finally, the images for each night were combined with a variance weighted mean (Supplementary Fig. 3). Before subtraction of artifacts, both pipelines deliver comparable performance.

**Data Availability:** All data from the NEAR campaign are publicly available at archive.eso.org under program ID 2102.C-5011(A). Original and processed data are also available from the corresponding author upon request.

**Acknowledgments:** NEAR was made possible by contributions from the Breakthrough Watch program, as well as contributions from the European Southern Observatory, including director's discretionary time. Breakthrough Watch is managed by the Breakthrough Initiatives, sponsored by the Breakthrough Prize Foundation. We would like to thank Rus Belikov, Eduardo Bendek,

Bernhard Brandl, Ryan Endsley, Rachel Fernandes, Kaitlin Kratter, Christian Marois, Michael Meyer, and Maxwell Moe for fruitful discussions and advice. The results reported herein benefited from collaborations and/or information exchange within NASA's Nexus for Exoplanet System Science (NExSS) research coordination network sponsored by NASA's Science Mission Directorate. This work was supported by the National Centre of Competence in Research PlanetS supported by the Swiss National Science Foundation, by the Fonds de la Recherche Scientifique - FNRS under Grant no F.4504.18, by the European Research Council (ERC) under the European Union's Horizon 2020 research and innovation program (grant agreement no 819155) and under the European Union's Seventh Framework Program (grant agreement no 337569), and by the Wallonia-Brussels Federation (grant for Concerted Research Actions). KW acknowledges support from NASA through the NASA Hubble Fellowship grant HST-HF2-51472.001-A awarded by the Space Telescope Science Institute, which is operated by the Association of Universities for Research in Astronomy, Incorporated, under NASA contract NAS5-26555. AB and SPQ acknowledge the financial support of the Swiss National Science Foundation. GR was supported by an NSF Astronomy and Astrophysics Postdoctoral Fellowship under award AST-1602444.

**Author contributions:** M. Kasper contributed as the NEAR experiment lead. K.W., A.B., P.P., M. Kasper, É. C., and S.P. Q. contributed to the data analysis. K.W., A.B., P.P., M. Kasper, and D. A. contributed to the preparation of the manuscript. K.W., A.B., U.K., A.-L. M., E.P., R.S., O.A., N.A., D.A., O.G., D.M., S.P. Q., M.S., A.V., and T.Z. contributed scientific advice. M. Kasper, U.K., A.-L. M, E.P., R.S., G.Z., O.A., and J.K. contributed to campaign observations. R.A. contributed to experiment management. G.J. contributed to experiment engineering (mechanics, system). U.K. contributed to experiment engineering (optics, system). A.-L. M., O.A., P.F., E.H., M. Karlsson, D.M., G. O.X., and G. R. contributed to experiment engineering (coronagraph). A.C. and G.R. contributed to coronagraph apodizer design. G.Z. and P.D. contributed to experiment engineering (software). N.A. N.A., S.G., D.K., S.L., A.R., and M.R. contributed to experiment engineering (system). C. D. contributed to experiment testing (coronagraph). K.D., M. N.D., and A.V. contributed to experiment engineering (non-common path aberrations). E.F. contributed to experiment engineering (cooling). O.G. and P. Klupar contributed to experiment oversight. P. Kervella contributed to the analysis of the background star hypothesis. J.-P. K. contributed to experiment engineering (mechanics). C.S. contributed to experiment engineering (electronics). J. K. contributed to adaptive optics operation. M. N.D. and M.R. contributed to experiment commissioning. M.S. and T.Z. contributed to project oversight.

**Competing interests:** The authors declare no competing interests.

# Supplementary Information for

## Imaging low-mass planets within the habitable zone of α Centauri

**Authors:** K. Wagner,[1, 2, *] A. Boehle,[3] P. Pathak,[4] M. Kasper,[4] R. Arsenault,[4] G. Jakob,[4] U. Käufl,[4] S. Leveratto,[4] A.-L. Maire,[5] E. Pantin,[6] R. Siebenmorgen,[4] G. Zins,[4] O. Absil,[5] N. Ageorges,[7] D. Apai,[1,2,8] A. Carlotti,[9] É. Choquet,[10] C. Delacroix,[5] K. Dohlen,[10] P. Duhoux,[4] P. Forsberg,[11] E. Fuenteseca,[4] S. Gutruf,[7] O. Guyon,[1, 12, 13] É. Huby,[15] D. Kampf,[7] M. Karlsson,[11] P. Kervella,[15] J.-P. Kirchbauer,[4] P. Klupar,[13] J. Kolb,[4] D. Mawet,[16] M. N'Diaye,[17] G. Orban de Xivry,[5] S. P. Quanz,[3] A. Reutlinger,[7] G. Ruane,[16, 18] M. Riquelme,[4] C. Soenke,[4] M. Sterzik,[4] A. Vigan,[10] and T. de Zeeuw[4, 19, 20]

[1] Dept. of Astronomy and Steward Observatory, University of Arizona, 933 N Cherry Ave, Tucson, AZ 85719, USA.

[2] NASA Nexus for Exoplanet System Science, Earths in Other Solar Systems Team.

[3] Institute for Particle Physics and Astrophysics, ETH Zurich, Wolfgang-Pauli-Strasse 27, 8093 Zürich, Switzerland.

[4] European Southern Observatory, Karl-Schwarzschild-Straße 2, 85748 Garching bei München, Germany.

[5] STAR Institute, Université de Liège, Allée du Six Août 19c, B-4000 Liège, Belgium.

[6] AIM, CEA, CNRS, Université Paris-Saclay, Université Paris Diderot, Sorbonne Paris Cité, F-91191 Gif-sur-Yvette, France.

[7] Kampf Telescope Optics, Alois-Gilg-Weg 7, 81373 München, Germany.

[8] Lunar and Planetary Laboratory, University of Arizona, 1629 E University Blvd, Tucson, AZ 85721, USA.

[9] Univ. Grenoble Alpes, CNRS, IPAG, 38000 Grenoble, France.

[10] Aix Marseille Univ, CNRS, CNES, LAM, Marseille, France.

[11] Department of Materials Science and Engineering, Ångström Laboratory, Uppsala University, Uppsala, Sweden.

[12] Subaru Telescope, 650 N Aohoku Pl, Hilo, HI 96720, USA.

[13] The Breakthrough Initiatives, NASA Research Park, Bld. 18, Moffett Field, CA, 94035, USA.

[14] James C. Wyant College of Optical Sciences, University of Arizona, Tucson, AZ, USA

[15] LESIA, Observatoire de Paris, 5 Place Jules Janssen, 92190 Meudon, France.

[16] California Institute of Technology, 1200 E. California Blvd., Pasadena, CA 91125, USA.

[17] Université Côte d'Azur, Observatoire de la Côte d'Azur, CNRS, Laboratoire Lagrange, Nice, France.

[18] Jet Propulsion Laboratory, California Institute of Technology, 4800 Oak Grove Dr., Pasadena, CA 91109, USA.

[19] Sterrewacht Leiden, Leiden University, Postbus 9513, 2300 RA Leiden, The Netherlands.

[20] Max Planck Institute for Extraterrestrial Physics, Giessenbachstraße 1, 85748 Garching, Germany.

* Correspondence to: kevinwagner@email.arizona.edu

**This PDF file includes:**

Supplementary Methods
Supplementary Figures 1 to 6
Supplementary Table 1
Supplementary References (*1–19*)

**Supplementary Methods**

Artifact Modeling I. Persistence Stripes

The most significant systematic artifact is detector persistence accumulated as a result of photons collected during the chopping sequence. These line-like features rotate with the field of view– resulting in reduced sensitivity throughout the entire image and greater false-positive potential in the areas directly between the stars. Either an empirical or a physically motivated model is needed to correct for this effect−similar to the charge trapping and delayed release models recently developed for HgCdTe detectors (e.g., *1, 2*). We constructed two persistence models, which are shown in Supplementary Fig. 1. The first is a simple empirical model that includes a superposition of lines convolved with the instrumental PSF. The second is a higher-fidelity model in which we simulated the chopping kinematics and charge accumulation. In this model, we assumed a persistence response that is proportional to the signal in the PSF and included a free parameter to scale the intensity of the persistence pattern of α Centauri B with respect to that of α Centauri A to remove the major contributions of non-linearity. We translated the PSFs of α Centauri A and B across a simulated image plane for both strokes of the chopping sequence, including a sub-pixel offset and corresponding re-centering behind the coronagraph at the end of each stroke. We do not account for a potential time decay of the signal and retain all accumulated intensity in the pattern. We subtracted the persistence model of the second stroke from that of the first, and then subtracted the combined model from the individual frames.

   To fit the parameters, we measured the positions of the off-axis PSFs of α Centauri A and B in each individual image and performed a grid search with velocity (degenerate with intensity, which we did not vary) and the offset from the center of the coronagraph as free parameters. We assumed that the offset for the second stroke is opposite that of the first to mimic a small error in the chopping angle. For each set of parameters within the grid, we constructed and subtracted the persistence model from each individual image prior to PSF subtraction, and repeated the remaining steps of the data reduction–primarily including high-pass filtering and PSF subtraction via the KLIP algorithm. We adopted the parameters that minimized the standard deviation of pixel values within the final combined image in an annulus from 10−45 pixels (∼0.45−2 arcsec).

   Both models are able to qualitatively reproduce the persistence pattern observed in the data (see Supplementary Fig. 1). Following model subtraction, a greater speckle density is present in the areas covered by persistence. The speckle intensity distribution in this region is clearly non-uniform, with several maxima present primarily within the persistence pattern from α Centauri A. Therefore, we refrain from characterizing sources in the areas impacted by persistence (between the dashed lines in Fig. 4). Additionally, the residual noise in the areas not impacted by persistence is improved following subtraction of the simulated model prior to PSF subtraction. This is because the persistence rotates with the field of view and biases the PSF subtraction. Accordingly, we utilize the images generated by the removal of the simulated persistence model prior to high-pass filtering and PSF subtraction for the proceeding analysis. The detection and properties of the candidate C1 are not affected by the choice in persistence model (including without subtracting any model).

Artifact Modeling II. Optical Ghosts

The second most significant artifact that appears in the final images are two dark arcs−likely bright features in the off-axis PSF of α Centauri A−that come within 1" of the coronagraphic residual. These features are likely faint optical ghosts that are introduced by the spectral filter, and which rotate with the parallactic angle to create arc-like features in the final image. The presence of these arcs weakens the overall image sensitivity. Similar to our handling of the detector persistence stripes, we created an empirical model of the dark arcs as circular segments of negative flux and subtracted this from the images before PSF subtraction. We repeated the process to optimize the parameters of the arcs: including radius, brightness, and width, while keeping the center fixed to the off-axis position of α Centauri A. Following the removal of the arcs, the image background within ~1" of the coronagraphic mask is nearly flat, with the exception of the persistence stripe residuals (see Supplementary Figs. 1–3).

## Plate Scale and True North Calibration

We utilized the orbit of the binary as reported in (3) to calculate the instrumental plate scale and true North offset (from the parallactic angle zero-point) to be used for astrometric calibrations. In June 2019, the orbit fit predicts that α Centauri B should appear at $\rho$=5.15 arcsec and $\theta$=−18.6º E of N with respect to α Centauri A. Based on this orbit fit and the apparent position of the stars in the median-combined, non-PSF subtracted images, we calculated a plate scale of 0.0456±0.0002 arcsec/pixel and a true North offset of −36.7º±0.1º East of North.

## Background-Limited Sensitivity

To compute the background-limited sensitivity, we measured the standard deviation of pixel values in a region of the detector far from α Centauri A and B, and compared this to the flux measured in a four-pixel radius aperture centered on the off-axis PSF of α Centauri A. We converted this ratio to a 3-σ sensitivity estimate by multiplying by a factor of $(3 \times \sqrt{16\pi})$ to account for the aperture area, and arrived at 5×10$^{-7}$ contrast with respect to α Centauri A, or about 65 μJy (the $N$-band flux of α Centauri A is 129 Jy, 4). The background near the stars is higher (by a factor of two at 1 arcsec) due to the increased photon noise and the glow of the AGPM (5).

## Simulated Planetary Contrast vs. Separation Curves

We first calculated the radiative equilibrium temperature of each simulated planet on a circular orbit as a function of its orbital semi-major axis ($a$), its Bond albedo ($A_B$), the stellar radius ($R$), and stellar temperature ($T_{star}$). This is given by Supplementary Equation (1), which is obtained by setting the planet's luminosity to be equivalent to its total input of irradiant energy from the star (6).

$$T_{\text{eq}} = T_{\text{star}} \sqrt{\frac{R}{2a}} (1 - A_{\text{B}})^{1/4} \quad (1)$$

We assumed $T_{\text{star}}$ = 5800 K and $R$ = 1.2 $R_{\odot}$ for A, and $T_{\text{star}}$ = 5250 K and $R$ = 0.86 $R_{\odot}$ for B (48). We then included additional heat as a fraction of the planet's equilibrium temperature via Supplementary Equation (2), and computed Blackbody spectra for the star and planet.

$$T_{\text{planet}} = T_{\text{eq}} * (1 + f_{\text{extra}}) \quad (2)$$

We then created simulated photometric measurements by integrating the spectra between 10–12.5 μm in steps of 0.1 μm. These synthetic photometric measurements were used to calculate planet-star contrast ratios. We repeated the analysis for planets of different radii, Bond albedo, orbital semi-major axes, and additional heating conditions to compute the simulated contrasts that are utilized to estimate the properties of C1 and also for comparisons in Figs. 1, 4, & 5.

Despite its relative simplicity, this approach is sufficient to gain an understanding of the sensitivity of our single-band photometric measurements. Specifically, uncertainties in planetary properties (e.g., molecular content and sources of additional heating) preclude the usefulness of a more sophisticated model. Our model does not include a wavelength-dependent model of molecular absorption and emission processes. Instead, absorption processes are incorporated in bulk via the Bond albedo. For Earth's atmosphere, the dominant absorbers in the $\lambda \sim 8–13$ µm range are $O_3$ at $\lambda < 10$ µm and $CO_2$ at $\lambda > 12.5$ µm (7). Therefore, the overall level of absorption for Earth-like planets is expected to be low within the $\lambda = 10–12.5$ µm range of the NEAR filter (8). For Neptune, the dominant feature within this range is $C_2H_6$ emission at 12.2 µm, which is an order of magnitude above the thermal continuum (9). For Jupiter, $NH_3$ absorption at ~11 µm is present at a comparable equivalent width to the also present $C_2H_6$ emission (10). Such planetary atmospheres at ~1 au would have comparable equilibrium temperatures to Earth, which will change the overall ratios of these constituents and their spectral contributions. However, more detailed modeling of such atmospheres also suggests that blackbody spectra are reasonable approximations for $\lambda = 10–12.5$ µm (11). This approach also does not account for specific processes that could be responsible for producing additional heating or non-thermal emission and includes the sum of all of these in a simple fractional parameter, $f_{\mathrm{extra}}$. If the solar system's giant planets are an accurate indication of the larger population, then these sources of additional heating can be expected to raise the thermal infrared planetary fluxes by up to an order of magnitude compared to the predictions of radiative thermal equilibrium (corresponding to $f_{\mathrm{extra}} \sim 0.8$, which is above the maximum of $f_{\mathrm{extra}} \sim 0.5$ considered here; see 12, 13).

While these models carry a significant uncertainty in the maximum luminosity, they establish useful lower limits in cases of low additional heat and Bond albedos representative of the solar system planets ($A_B \sim 0.3$ for Earth, 6). Finally, we verified the predicted contrasts by comparing to the Earth's brightness at $\lambda \sim 10$ µm (7). A planet around α Centauri A would experience a similar irradiance as the Earth if it were at a distance of ~1.2 au, or ~0.9 arcsec. Scaled to a radius of 1.7 $R_\oplus$ and shifted to the distance of α Centauri (1.3 pc), such a planet would appear with a contrast of $\sim 5 \times 10^{-7}$. Our corresponding model of a super-Earth with $f_{\mathrm{extra}} = 0.1$, with the same planetary radius (1.7 $R_\oplus$), and at the same separation from the star (0.9 arcsec) has a contrast of $\sim 4 \times 10^{-7}$. This serves as a useful consistency check of our models and assumptions.

Simulated Planet Injection and Retrieval Tests
To compute the image sensitivity, we performed extensive point source injection and recovery tests (see Figs. 3 & 4). For the injected PSF, we utilized the off-axis PSF of α Centauri A and included a scaling correction for the transmission of the AGPM (14). The PSFs were injected prior to high-pass filtering, after which we completed the remaining data processing steps (PSF subtraction, etc.). We measured the signal within an aperture of two pixels in diameter centered on the location of the injected source. This diameter corresponds to ~0.3×FWHM, which is smaller than typically used. This provides a larger number of apertures that are non-contaminated by persistence. To measure the noise, we blocked the regions of the image affected by detector persistence and a 12-pixel wide box centered on the injected source and calculated the mean and standard deviation of fluxes within all non-overlapping apertures at the same separation. We converted these measurements into a signal-to-noise ratio (SNR) estimate (*15*), adopting SNR=3 as a detection threshold, as this resulted in all of the automatically retrieved sources also being identified by eye in the images (see Fig. 3). We caution that as a result of the residual structures in the images, the statistics are non-Gaussian, and probabilities derived from Gaussian statistics should not be applied to these measurements. We repeated the analysis with injected sources along ten different position angles and from separations of 0.3 to 1.5 arcsec. We adopted the average sensitivity level over the tested position angles as the sensitivity for each separation in the contrast curves shown in Fig. 4. For the completion analysis (see Fig. 5 and following subsection), we adopted the maximum brightness of the recovered sources as the sensitivity for each separation.

## Completeness Analysis

We converted the sensitivity analysis into completion estimates using a Monte Carlo simulation to draw randomly sampled orbital parameters. For different sets of uniformly sampled planet radius and semi-major axis, we sampled the argument of periapsis ($\omega \propto$ unif[0 , 360]; this parameter is degenerate with the time of closest approach for the circular orbits considered here), angle of ascending node ($\Omega \propto$ unif[0 , 360]), and inclination ($i \propto \sin i$) and then determined the planet's separation and position angle on the sky. The radiative equilibrium temperature was determined assuming a Bond albedo of 0.3 and then increased by 10% and 50% to account for different amounts of internal/additional heat from the planet. The temperature was converted to an *N*-band contrast following the previous subsection. This contrast was then compared to the maximum point-source brightness at that separation that would have been detected based on simulated point-source injections (see above). The completeness maps are shown in Fig. 5. We also tested restricting the plane of the planetary orbits to the orbital plane of the binary, and also allowed for eccentric orbits, and found that these did not significantly alter the completeness maps.

## Properties of The Candidate Detection 'C1'

Based on a similar analysis to the simulated planet injection and recovery tests, C1 is detected with *SNR*~3.1 (using apertures with diameter of two pixels, resulting in 27 noise measurements not contaminated by the persistence artifact). Using apertures with diameter of 1×FWHM, corresponding closely to the definition of *SNR* in (*15*), gives *SNR*~3.5, although with only ten independent noise measurements after rejecting seven due to persistence contamination. Again, we caution that probabilities derived from Gaussian statistics should not be applied to this *SNR* estimate. To measure the brightness and position of C1, we performed negative point source injections over a grid in radius, position angle, and contrast to minimize the residual squared flux in the total image area (see the bottom-left panel of Supplementary Fig. 4). We injected the point sources in the data processing prior to high-pass filtering and PSF subtraction and performed a similar analysis with both pipelines. The findings were consistent, and we averaged the results to arrive at $\rho = 0.85 \pm 0.05$ arcsec, $\theta$=228.9º ±3.3º E of N, and $F_{C1}/F_{\alpha CenA} = 9.3 \times 10^{-6} \pm 3.1 \times 10^{-6}$, corresponding to $F_{C1}$=1.2±0.4 mJy, after correcting for the ~20% throughput of the coronagraph, and by estimating the astrometric uncertainties as the FWHM divided by the SNR. Assuming that the emission comes from a planet with a Bond albedo of ≤0.3 (similar to Earth), and that the planet's temperature is between 100−150% of its radiative equilibrium temperature ($T_{eq}$~300 K) to account for the possibility of additional sources of heat (such as greenhouse heating or residual heat from the planet's formation), and assuming blackbody spectra, this brightness and projected separation correspond to a radius of ~3.3−11 R$_\oplus$. This is a plausible radius and level of additional heating compared to the ice− and gas-giants within our own solar system (e.g., *12, 13*).

## Pre-imaging for Background Sources

We first ruled out the possibility that C1 may be a background source by utilizing $K$-band ($\sim 2$ μm) images taken in 2009 (*3*). The faintest source identifiable in these earlier images is a $K \sim 13.6$ star that approaches the trajectory of α Centauri B in 2021. There is nothing comparable in brightness and proximity in mid-2019. Given that the NEAR campaign images are sensitive to $N \sim 12−12.5$, and that even the reddest stars have neutral mid-to-near infrared colors, we can exclude the possibility that C1 is a background star. If C1 is a background galaxy, it would likely have a rest peak wavelength of $\lambda \sim 1$ μm and would thus be redshifted by $Z \sim 10$. Such a galaxy would be among the most distant galaxies observed, and unlikely to appear projected on the sky within an arcsecond of α Centauri A and with such an apparent brightness. In all of the HST legacy fields spanning >800 arcmin$^2$, only a handful of such galaxies were detected (*16*), and all have near-IR magnitudes >25, which is much fainter than our observations would have been able to detect.

## Exozodiacal Dust Disk Modeling

We simulated an exozodiacal dust disk at the central observing wavelength of $\sim 11.25$ μm using *zodipic* (*17*). The code enables simulating both the disk and the star simultaneously, which, when convolved to the resolution of the observations, enables straightforward model flux calibration. We assumed a 79º inclination to be coplanar to the orbit of the binary (*3*) and fit the position angle and small central offset. Finally, we ran a grid search over the amount of dust in the model disk to match the brightness of C1. We subtracted each model from the data prior to high-pass filtering and PSF subtraction and adopted the model that minimized the squared residuals in an aperture of 10−45 pixels from the center as the best-fit. The results are shown in Supplementary Fig. 4 in comparison with the residuals from the point source subtraction.

## Astrometry and Orbital Modeling

To measure the astrometry of C1, we combined back-to-back nights, and re-performed minimization of the image residuals by negative synthetic planet injection tests. The astrometric precision was estimated as 1/3 of a beam diameter, or $\sim 8$ mas. The measurements are shown in Supplementary Fig. 5. To check consistency of the measurements with orbital motion, we generated 10,000 sample orbits using the OFTI method (*18*) in the orbitize! package (*19*) with the default settings (i.e., uniform priors). One hundred randomly selected sample orbits are shown in Supplementary Fig. 5 along with the retrieved orbital parameters. The data are consistent with a co-planar and dynamically stable orbit interior to the α Centauri AB binary ($a$<3 au and $i \sim 70º$). While the astrometric data are consistent with a Keplerian orbit, we caution that this does not constitute proof of orbital motion, as the data are also consistent with no motion.

The measurements show a tentative change of $\Delta\rho \sim −0$ .1 arcsec (i.e., toward the star) and $\Delta$PA $\sim 8º$ from 2019-05-24 to 2019-06-11, which establishes a useful vector to check against the data taken in late June. Projected forward in time, C1 would likely be at a separation of $\rho \sim 0.6−0.7$ arcsec on 2019 June 27. At this position, C1 would overlap the area of the image that is strongly impacted by the persistence stripes, and thus would not likely be detected. While C1 is not detected on this night, forward modelling of the anticipated signal assuming zero motion suggests that such a source would likely have been detected (see Supplementary Fig. 6). The non-detection of C1 on June 27, compared to its identification in $\sim 85\%$ of the single-night images taken prior to this date, tentatively supports the planetary hypothesis over those invoking exozodiacal dust or an unknown instrumental artifact, which would remain in a stable position throughout the campaign.

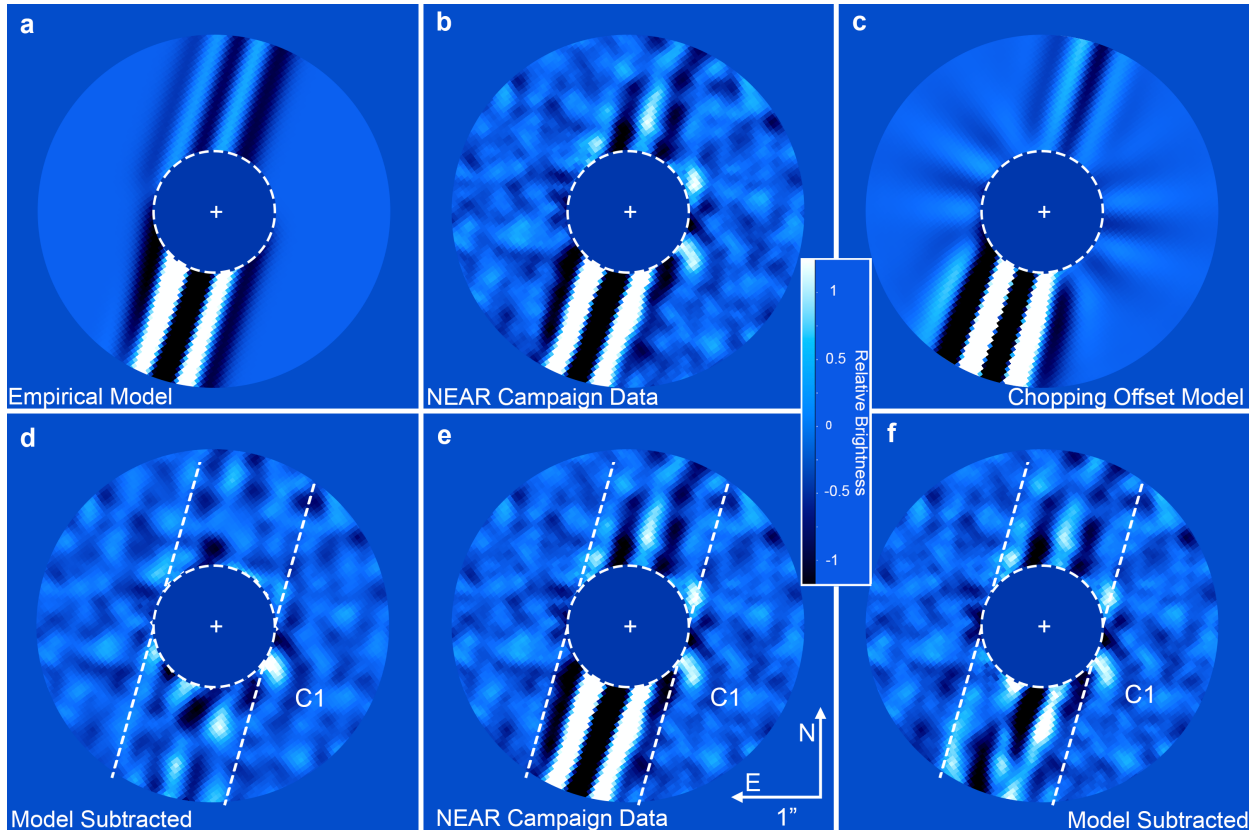

**Supplementary Fig. 1. Detector persistence models and the NEAR campaign image following correction.** The empirical model fit to the post-KLIP image (**a** and **d**) is subtracted after the KLIP procedure, while the simulated chopping offset model (**c** and **f**) is subtracted prior to the KLIP procedure. The central panels **b** and **e** show the original data without persistence model subtraction for comparison. The model image in **c** has been processed with KLIP to enable a straightforward comparison to **b** and **e**. Areas impacted by persistence are marked with dashed lines. We only consider sources exterior to these regions as potentially real detections. A point-like feature (C1) is marked as a candidate detection. Artifacts from optical ghosts have also been removed.

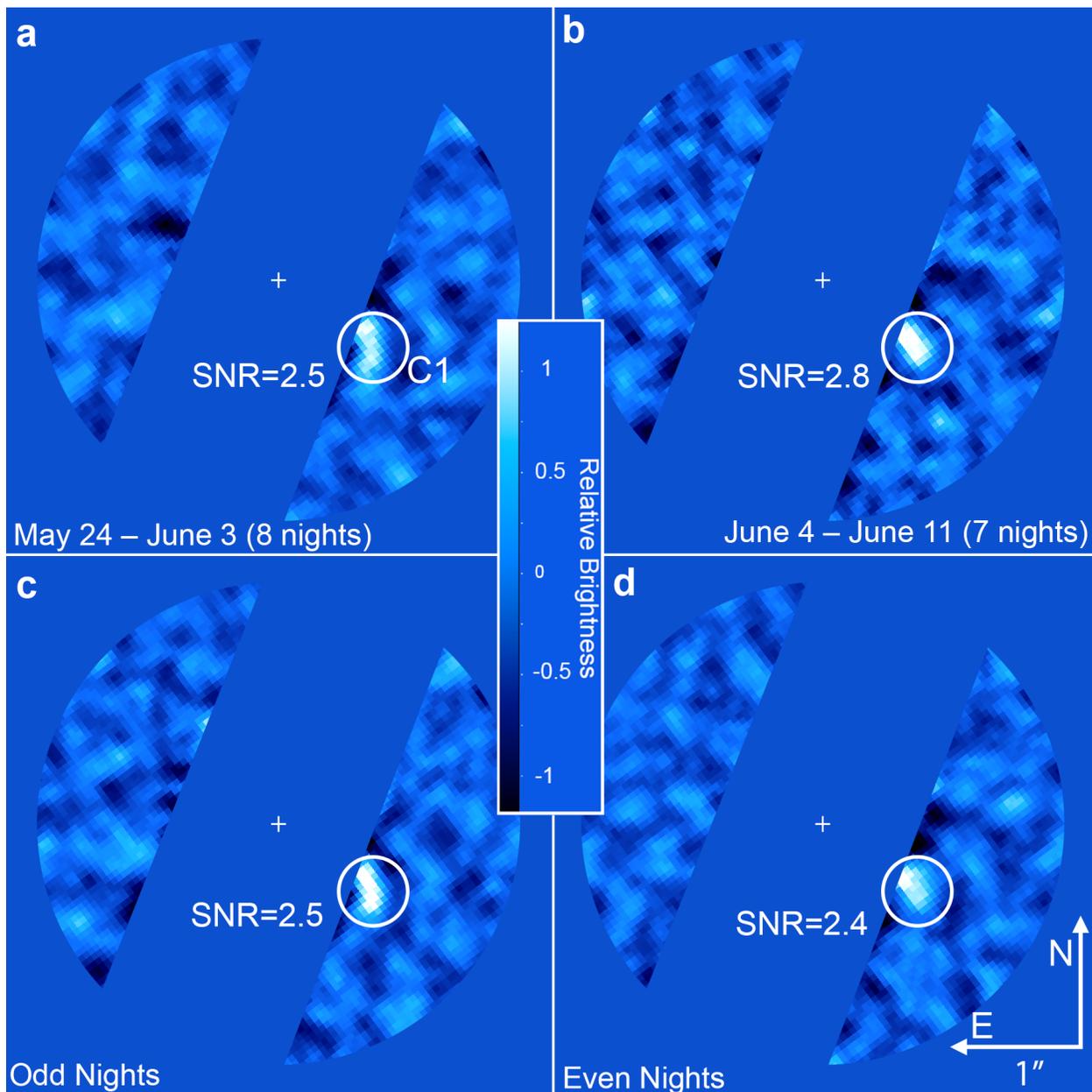

**Supplementary Fig. 2. NEAR campaign data split into two sets of independent datasets. a** & **b** compare the first and second halves of the campaign, respectively, while **c** & **d** compare even and odd nights. C1 is the only source clearly identified in each image outside of the area impacted by the persistence stripes (masked here).

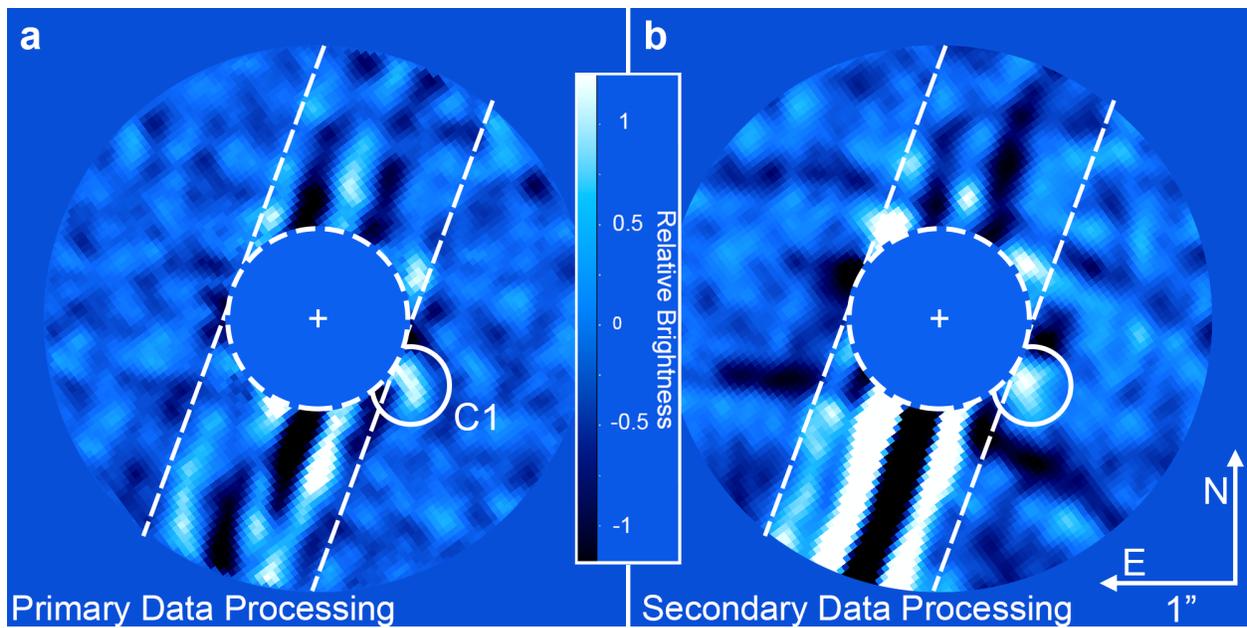

**Supplementary Fig. 3. NEAR data processed by two completely independent data processing pipelines.** The primary data processing (**a**) implements removal of the known systematic artifacts, whereas the persistence stripes and dark arcs are present in **b**. Excluding the known instrumental artifacts, C1 is the only source with a repeated detection in both images.

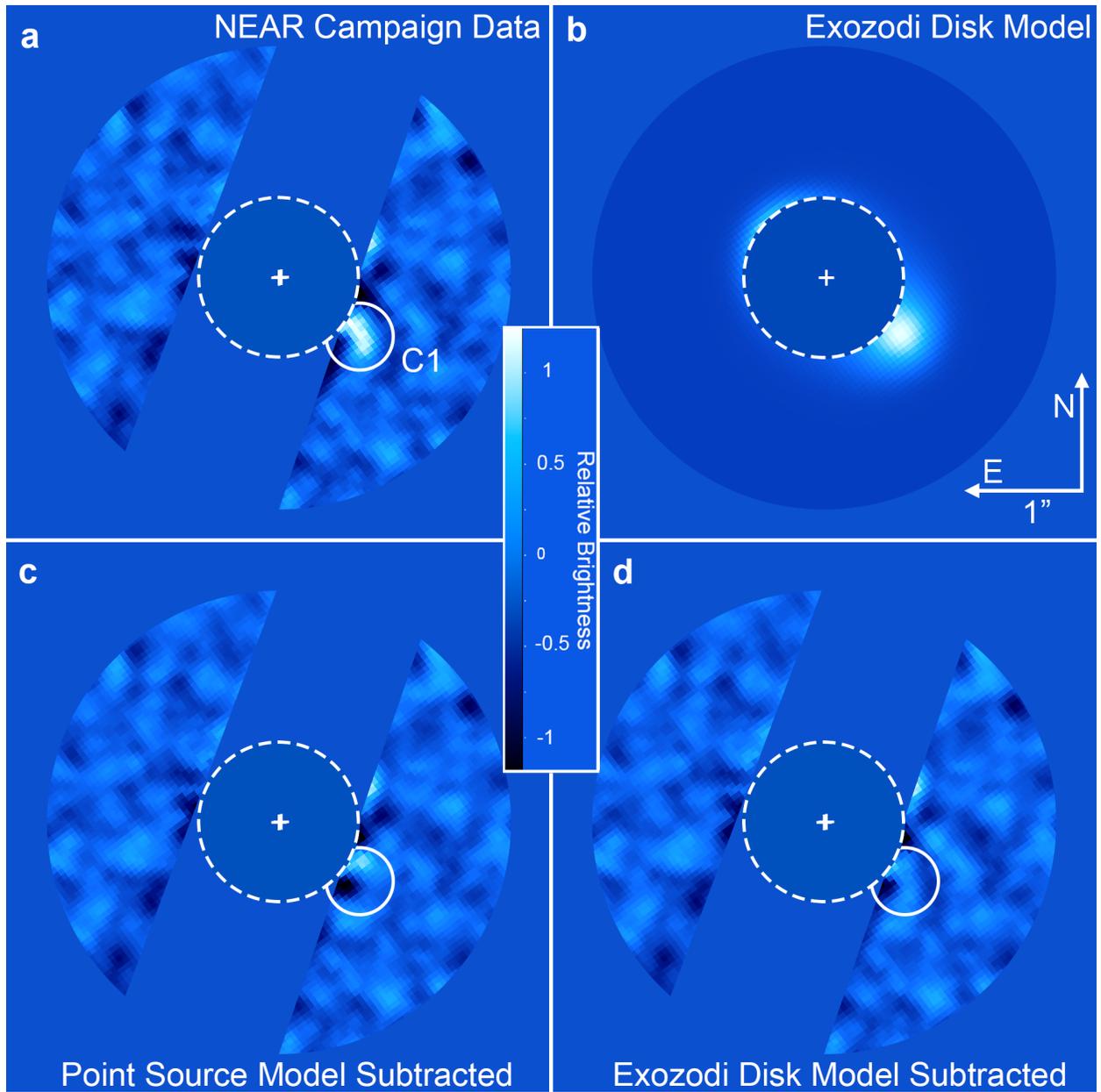

**Supplementary Fig. 4. C1 subtracted via both an exoplanet and exozodiacal disk models. a** shows the data without C1 subtracted for comparison. **c** shows the subtraction of C1 with a point source (i.e. exoplanet) model. **b** shows the model disk, which is inclined by 79º from the plane of the sky, extends from 0.8−1.2 au, includes 60 zodis of dust, and is offset by ~0.3 au in the direction of C1. **d** shows the image with the disk subtracted. Both models match the data reasonably well.

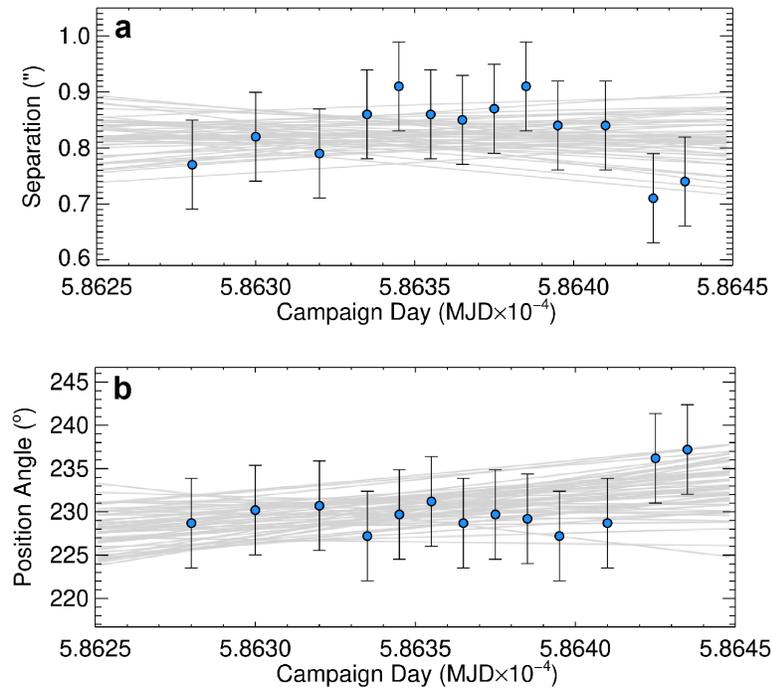

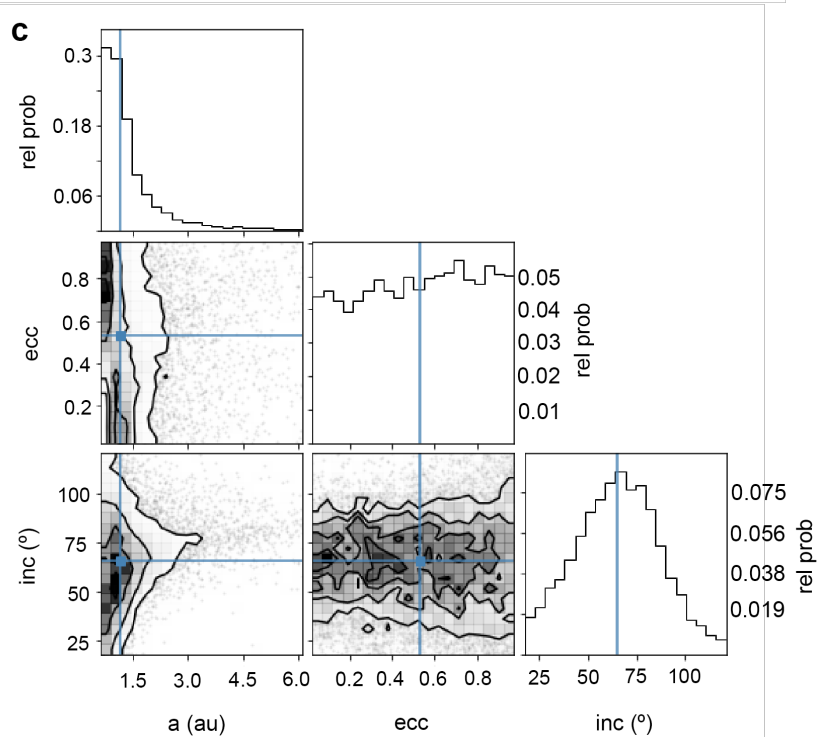

**Supplementary Fig. 5. Astrometric data and orbital fit for C1 generated from the NEAR campaign.** The astrometric data (**a, b**) are consistent with a bound orbit with a semi-major axis of ~1.1 au and an inclination of 65º±25º (**c**). Error bars in **a** and **b** represent the estimated astrometric uncertainty of 1/3 of a beam diameter, or ~8 mas. While the nature of the candidate is not yet certain, the consistency of a Keplerian orbit and its alignment with the orbit of the binary is a useful check for the hypothesis that C1 is a planet orbiting α Centauri A.

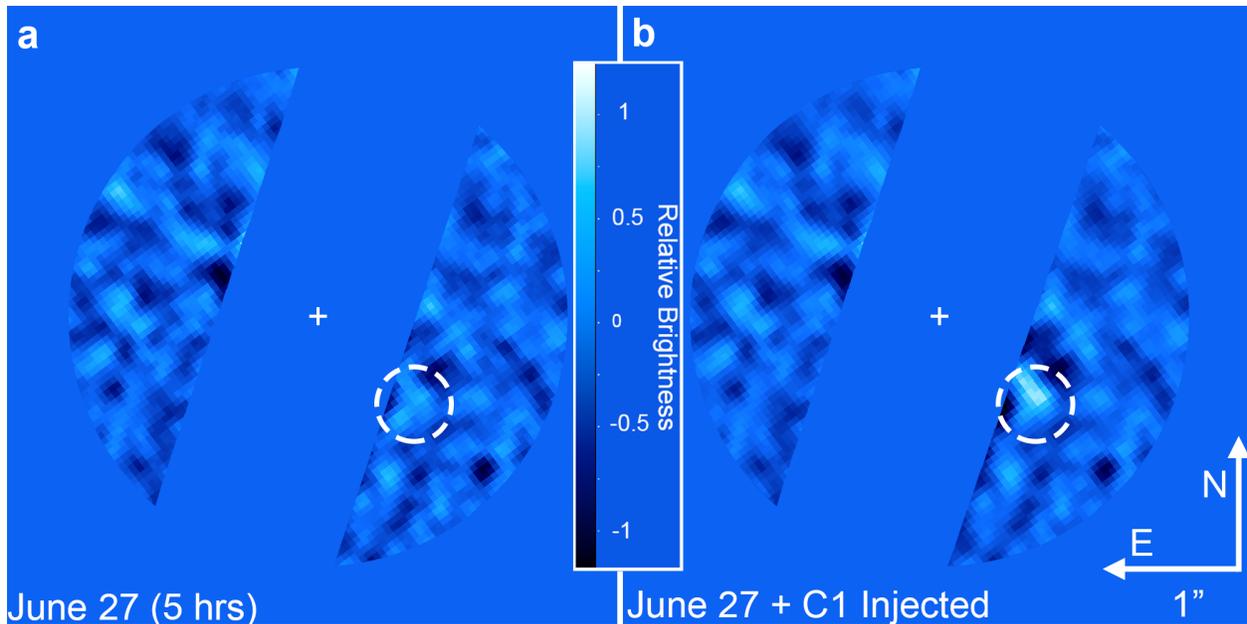

**Supplementary Fig. 6**. **Observations of α Centauri with NEAR on 2019 June 27.** The approximate position of C1 is indicated by the dashed circle. **a** shows the original data, and panel **b** shows the same data with a simulated planet with the properties of C1. These images illustrate that C1 would have likely resulted in a detection on June 27 if it remained stable in position and brightness.

| UT Date 2019- | Int. Time (hr) | Avg. Seeing (") | Ambient Temp (°C) | PWV (mm) |
|---|---|---|---|---|
| 5-24 | 6.3 | 1.0 | 10–12 | <1 |
| 5-26 | 5.8 | 1.2 | 8–10 | 2.5–5 |
| 5-28 | 2.4 | 1.0 | 7–10 | 5 |
| 5-29 | 0.8 | 1.0 | 9–12 | 7.5 |
| 5-30 | 5.9 | 0.7 | 9–10 | 6.5 |
| 5-31 | 5.5 | 1.0 | 7–10 | 3.0 |
| 6-01 | 5.3 | 1.0 | 7–10 | 2.5–4 |
| 6-02 | 4.6 | 1.0 | 10 | <1 |
| 6-03 | 6.4 | 0.7 | 15 | 2–2.5 |
| 6-04 | 6.3 | 0.7 | 17 | 2.5–3 |
| 6-05 | 5.4 | 1.0 | 17 | 1–2.5 |
| 6-06 | 3.3 | 1.2 | 14–16 | 1.25 |
| 6-08 | 2.7 | 0.5 | 12–17 | 1.1 |
| 6-09 | 5.9 | 0.7 | 15 | 1.25–2 |
| 6-10 | 3.1 | 0.5 | 12–17 | 2–2.25 |
| 6-11 | 2.2 | 1.5 | 12–14 | 2.5–3 |
| 6-27 | 5.0 | 1.2 | 9–10 | 2.5–5 |
| | Total=76.9 | Med.=1.0 | Med.=11 | Med.=2.8 |

**Supplementary Table 1. Near Campaign Observing Log.** Note: PWV = Precipitable Water Vapor. Seeing values correspond to differential image motion monitor (DIMM) measurements at $\lambda \sim 0.5$ μm.

**Supplementary References**: